\documentclass{article}

\usepackage{nmr}

\usepackage{latexsym}
\usepackage[matrix,arrow]{xy}
\CompileMatrices

\newtheorem{theorem}{Theorem}
\newtheorem{proposition}[theorem]{Proposition}

\newtheorem{corollary}[theorem]{Corollary}
\newtheorem{definition}{Definition}
\newtheorem{example}{Example}
\newtheorem{remark}{Remark}

\newenvironment{proof}[1][Proof.]{\noindent\textbf{#1} }{\hfill$\Box$}

\newcommand{\herbrand}{{\cal H}}
\newcommand{\assm}{{\it assm}}
\newcommand{\conc}{{\it conc}}
\newcommand{\mycenter}[1]{\hspace*{\fill}#1\hspace*{\fill}}

\author{Ralf Schweimeier and Michael Schroeder\\
Department of Computing\\ 
City University, London, UK\\
\{ralf,msch\}@soi.city.ac.uk}
\title{Well-Founded Argumentation Semantics for Extended Logic Programming}

\begin{document}

\maketitle

\begin{abstract}
  This paper defines an argumentation semantics for extended logic
  programming and shows its equivalence to the well-founded semantics
  with explicit negation.  We set up a general framework in which we
  extensively compare this semantics to other argumentation semantics,
  including those of Dung, and Prakken and Sartor.  We present a
  general dialectical proof theory for these argumentation semantics.
\end{abstract}

\section{Introduction}

Argumentation has attracted much interest in the area of AI. On the
one hand, argumentation is an important way of human interaction and
reasoning, and is therefore of interest for research into intelligent
systems. Application areas include automated negotiation via
argumentation~\cite{PSJ98:AgentsArguing,KSE98:Argument,Sch99:Argument}
and legal reasoning~\cite{PS97:Argument}.  On the other hand,
argumentation provides a formal model for various assumption based (or
non-monotonic, or default) reasoning
formalisms~\cite{BDKT97:Argument,CML00:LogicalModelsOfArgument}.  In
particular, various argumentation based semantics have been proposed
for logic programming with default
negation~\cite{BDKT97:Argument,Dun95:Argument}.

Argumentation semantics are elegant since they can be captured in an
abstract
framework~\cite{Dun95:Argument,BDKT97:Argument,%
Vre97:AbstractArgumentationSystems,JV99:RobustSemantics},
for which an elegant theory of attack, defence, acceptability, and
other notions can be developed, without recourse to the concrete
instance of the reasoning formalism at hand. This framework can then
be instantiated to various assumption based reasoning formalisms.
Similarly, a dialectical proof theory, based on dialogue trees,
can be defined for an abstract
argumentation framework, and then applied to any instance of such a
framework~\cite{Dun95:Argument,JV99:DialecticSemantics}.

In general, an argument $A$ is a proof which may use a set of
defeasible assumptions.  Another argument $B$ may have a conclusion
which contradicts the assumptions or the conclusions of $A$, and
thereby $B$ {\em attacks} $A$.  There are two fundamental notions of such
attacks: undercut and rebut~\cite{PS97:Argument} or equivalently {\em
ground-attack} and {\em reductio-ad-absurdum attack}
\cite{Dun93:ArgumentExplicit}.  We will use the terminology of
undercuts and rebuts. Both attacks differ in that an undercut attacks a
premise of an argument, while a rebut attacks a
conclusion. 

Given a logic program we can define an argumentation semantics by
iteratively collecting those arguments which are acceptable to a
proponent, i.e. they can be defended against all opponent attacks.  In
fact, such a notion of acceptability can be defined in a number of
ways depending on which attacks we allow the proponent and opponent to use.

Normal logic programs do not have negative conclusions, which means
that we cannot use rebuts. Thus both opponents can only launch
undercuts on each other's assumptions.
Various argumentation semantics have been defined for normal logic programs
\cite{BDKT97:Argument,Dun95:Argument,KT99:ComputingArgumentation}, 
some of which are equivalent to 
existing semantics such as the stable model semantics~\cite{GL88:StableModels}
or the well-founded semantics~\cite{GRS91:WFS}.

Extended logic programs
\cite{GL90:ClassicalNegation,AP96:WFSX,wag94a}, on the other hand,
introduce explicit negation, which states that a literal is explicitly
false. As a result, both undercuts and rebuts are possible forms of attack;
there are further variations depending on whether any kind of counter-attack
is admitted. A variety of argumentation semantics arise if one allows
one notion of attack as defence for the proponent, and another as attack
for the opponent.
Various argumentation semantics have been proposed for extended logic programs
\cite{Dun93:ArgumentExplicit,PS97:Argument}. 
Dung has shown that a certain argumentation semantics is equivalent to
the answer set semantics~\cite{GL90:ClassicalNegation}, a generalisation
of the stable model semantics~\cite{GL88:StableModels}.
To our knowledge, no argumentation semantics has yet been found equivalent
to the well-founded semantics for extended logic programs, 
WFSX~\cite{PA92:WFSX,AP96:WFSX}.

This paper makes the following contributions: we define a least
fixpoint argumentation semantics for extended logic programs, and show
its equivalence to the well-founded semantics with explicit
negation~\cite{PA92:WFSX,AP96:WFSX,ADP95:LPsystem}.  In order to
relate this semantics to other argumentation semantics, we set up a
general framework to classify notions of justified arguments, and use
it to compare our argumentation semantics to those of
Dung~\cite{Dun93:ArgumentExplicit} and Prakken and
Sartor~\cite{PS97:Argument} among others.  We develop a general
dialectical proof theory for the notions of justified arguments we
introduce.

The paper is organised as follows: First we define arguments and
notions of attack and acceptability. Then we set up a framework for
classifying different least fixpoint argumentation semantics,
based on different notions of attack. In Section~\ref{sec:wfsx}, we recall
the definition of WFSX, and in Section~\ref{sec:wfsxarg}, we prove
the equivalence of an argumentation semantics and WFSX.
A general dialectical proof theory for arguments is presented in
Section~\ref{sec:proof-theory}, and its soundness and completeness is
proven.

\section{Extended Logic Programming and Argumentation}
\label{sec:elp}

We summarise the definitions of arguments for extended logic programs,
and define various notions of attack between arguments.

\subsection{Arguments}
\label{subsec:arguments}

\begin{definition} 
  An {\em objective literal} is an atom $A$ or its explicit negation
  $\neg A$. We define $\neg \neg L = L$.
  A {\em default literal} is of the form $not~L$ where $L$
  is an objective literal. A {\em literal} is either an objective or a
  default literal.  \\
  An {\em extended logic program} is a (possibly infinite) set of rules of
  the form \vspace{-1ex}
  $$L_0 \gets L_1,\dots,L_m, not~L_{m+1},\dots, not~L_{m+n}\vspace{-1ex}$$
  where $m, n \ge 0$, and each $L_i$ is an objective literal 
  $(0\leq i \leq m+n)$. \\
  For such a rule $r$, we call $L_0$ the {\em head} of the rule, 
  $head(r)$, and $L_1,\ldots,not~L_{m+n}$ the {\em body} of the rule, 
  $body(r)$.
\end{definition}

Our definition of an argument for an extended logic program is based
on~\cite{PS97:Argument}. 
Essentially, an argument is a partial proof, resting on a number of
{\em assumptions}, i.e.~a set of default literals.%
\footnote{In~\cite{BDKT97:Argument,Dun93:ArgumentExplicit}, an argument 
  {\em is} a set of assumptions; the two approaches are equivalent in 
  that there is an argument with a conclusion $L$ iff there is a set
  of assumptions from which $L$ can be inferred. See the discussion
  in~\cite{PS97:Argument}.}

Note that we do not consider priorities of arguments, as used e.g.\
in~\cite{PS97:Argument,Vre97:AbstractArgumentationSystems}.

\begin{definition} 
Let $P$ be an extended logic program. \\
An {\em argument} for $P$ is a finite sequence 
$A=[r_1,\dots r_n]$ of rules $r_i \in P$ such that
for every $1\le i \le n$, for every objective literal $L_j$ in
the body of $r_i$ there is a $k>i$ such that $head(r_k) = L_j$. \\
A {\em subargument} of $A$ is a subsequence of $A$ which is an argument.
The head of a rule in $A$ is called a {\em conclusion} of $A$,
and a default literal $not~L$ in the body of a rule of $A$ is called an
{\em assumption} of $A$.
We write $\assm(A)$ for the set of assumptions and $\conc(A)$ for the 
set of conclusions of an argument $A$. \\
An argument $A$ with a conclusion $L$ is a {\em minimal argument for $L$} 
if there is no subargument of $A$ with conclusion $L$.
An argument is {\em minimal} if it minimal for some literal $L$.
Given an extended logic program $P$, we denote the set of minimal arguments
for $P$ by $\mathit{Args}_P$.

\end{definition}

The restriction to minimal arguments is not essential, but convenient,
since it rules out arguments constructed from several unrelated
arguments. Generally, one is interested in the conclusions of an
argument, and wants to avoid having rules in an argument which do not
contribute to the desired conclusion.

\subsection{Notions of Attack}
\label{subsec:attack}

There are two fundamental notions of attack: {\em undercut},
which invalidates an assumption of an argument, and
{\em rebut}, which contradicts a conclusion of an 
argument~\cite{Dun93:ArgumentExplicit,PS97:Argument}.
From these, we may define further
notions of attack, by allowing either of the two fundamental kinds of
attack, and considering whether any kind of counter-attack is allowed
or not. We will now formally define these notions of attacks.

\begin{definition}
  \label{def:notionsofattack}
  Let $A_1$ and $A_2$ be arguments.
  \begin{enumerate}

  \item 
    $A_1$ {\em undercuts} $A_2$ if there is an objective literal
    $L$ such that $L$ is a conclusion of $A_1$ and $not~L$ is an 
    assumption of $A_2$.

  \item 
    $A_1$ {\em rebuts} $A_2$ if there is an objective literal $L$
    such that $L$ is a conclusion of $A_1$ and $\neg L$ is a
    conclusion of $A_2$.

  \item 
    $A_1$ {\em attacks} $A_2$ if $A_1$ undercuts or rebuts $A_2$.

\item 
  $A_1$ {\em defeats} $A_2$ if 
     $A_1$ undercuts $A_2$, or
     ($A_1$ rebuts $A_2$ and $A_2$ does not undercut $A_1$).
    
  \item 
    $A_1$ {\em strongly attacks} $A_2$ if $A_1$ attacks $A_2$ and
    $A_2$ does not undercut $A_1$.

  \item 
    $A_1$ {\em strongly undercuts} $A_2$ if $A_1$ undercuts $A_2$
    and $A_2$ does not undercut~$A_1$.
  \end{enumerate}
\end{definition}

The notions of \emph{undercut} and \emph{rebut}, and hence \emph{attack} are
fundamental for extended logic
programs~\cite{Dun93:ArgumentExplicit,PS97:Argument}. The notion of
\emph{defeat} is used in~\cite{PS97:Argument}, along with a notion of
\emph{strict defeat}, i.e.\ a defeat that is not counter-defeated.
For arguments without priorities, rebuts are symmetrical, and
therefore strict defeat coincides with strict
undercut, i.e.\ an undercut that is not counter-undercut.
Similarly, strict attack coincides with strict undercut.
For this reason, we use the term {\em strong undercut} instead of 
{\em strict undercut}, and similarly define {\em strong attack} to be
an attack which is not counter-undercut.
We will use the following abbreviations for these notions of attack.
\textit{r} for rebuts, 
\textit{u} for undercuts, 
\textit{a} for attacks, 
\textit{d} for defeats, 
\textit{sa} for strongly attacks, and
\textit{su} for strongly undercuts.

These notions of attack define for any extended logic program a binary
relation on the set of arguments of that program.

\begin{definition}
  A {\em notion of attack} is a function $x$ which assigns to each
  extended logic program $P$ a binary relation $x_P$ on the set of
  arguments of $P$, i.e.\ $x_P \subseteq \mathit{Args}_P^2$.  Notions
  of attack can be partially ordered by defining
  $x \subseteq y \mbox{ ~iff~ } \forall P: x_P \subseteq y_P$
\end{definition}

\begin{definition}
  Let $x$ be a notion of attack. Then 
  the {\em inverse} of $x$, denoted by $x^{-1}$, is defined as
  $x^{-1}_P = \{ (B,A) ~|~ (A,B) \in x_P \}$.
\end{definition}

In this relational notation,
Definition~\ref{def:notionsofattack} can be rewritten as 
$a = u \cup r$, $d = u \cup (r - u^{-1})$,
$sa = (u \cup r) - u^{-1}$, and $su = u - u^{-1}$.
Using the set-theoretic laws $A-B \subseteq A \subseteq A \cup C$ and
$(A \cup B) - C = (A - C) \cup (B - C)$ (for all sets $A$, $B$, and $C$),
it is easy to see that the notions of attack of
Definition~\ref{def:notionsofattack} are partially ordered according
to the following Hasse diagram.
\noindent
$$\hspace{-1.5cm}\xymatrixnocompile@C=-5em@R=1em{
& \parbox{3.5cm}{\begin{center}\textit{attacks} $= a = u \cup r$\end{center}\vspace{-1ex}} \ar@{-}[d] \\
& \parbox{4cm}{\vspace{-1ex}\begin{center}\textit{defeats} $=$ $d = u \cup (r - u^{-1})$\end{center}\vspace{-1ex}} 
\ar@{-}[dl] \ar@{-}[dr] \\
\hspace{1cm}\parbox{3.5cm}{\vspace{-1ex}\begin{center}\textit{undercuts} $= u$\end{center}\vspace{-1ex}} \ar@{-}[dr] && 
\hspace{-2cm}\parbox{5.5cm}{\vspace{-1ex}\begin{center}\textit{strongly attacks} $=$ $sa = (u \cup r) - u^{-1}$\end{center}\vspace{-1ex}} 
\ar@{-}[dl] \\
& \parbox{5.5cm}{\hspace{-2cm}\vspace{-1ex}\begin{center}\textit{strongly undercuts} $= su = u - u^{-1}$\end{center}} \\
}$$

This diagram contains the notions of attack used
in~\cite{Dun93:ArgumentExplicit,PS97:Argument}, plus
\textit{strongly attacks} which seemed a natural intermediate 
notion between \textit{strongly undercuts} and \textit{defeats}.
We have not included \textit{rebuts}, because in the absence of
priorities, \textit{rebuts} is somewhat weaker than \textit{undercuts}, 
because it is symmetric: a rebut is always counter-rebutted, 
while the same does not hold for \textit{undercuts}.

\subsection{Acceptability and Justified Arguments}
\label{subsec:acceptable}

Given the above notions of attack, we define acceptability of an
argument. Basically, an argument is acceptable if it can be defended
against any attack.  Depending on which particular notion of attack we
use as defence and which for the opponent's attacks, we obtain a host
of acceptability notions.

Acceptability forms  the basis for our argumentation semantics,
which is defined as the least fixpoint of a function, which collects
all acceptable arguments.  The {\em least} fixpoint is of particular
interest \cite{PS97:Argument,Dun93:ArgumentExplicit}, because it
provides a canonical fixpoint semantics and it can be constructed
inductively.

\begin{definition}
  Let $x$ and $y$ be notions of attack. Let $A$ be an argument, and
  $S$ a set of arguments. Then $A$ is {\em $x/y$-acceptable wrt.\ $S$}
  if for every argument $B$ such that $(B,A) \in x$ there exists an
  argument $C \in S$ such that $(C,B) \in y$.
\end{definition}

Based on the notion of acceptability, we can then define a fixpoint
semantics for arguments.

\begin{definition}
  Let $x$ and $y$ be notions of attack, and $P$ an extended logic
  program. The operator 
  $F_{P, x/y}:{\cal P}(\mathit{Args}_P) \rightarrow {\cal P}(\mathit{Args}_P)$
  is defined as
  $$F_{P, x/y}(S) = \{ A ~|~ A \mbox{ is $x/y$-acceptable wrt.\ $S$} \}$$
  We denote the least fixpoint of $F_{P, x/y}$ by $J_{P, x/y}$.
  If the program $P$ is clear from the context, we omit the subscript $P$.
  An argument $A$ is called $x/y$-justified if $A \in J_{x/y}$;
  an argument is called $x/y$-overruled if it is attacked by an
  $x/y$-justified argument; and an argument is called $x/y$-defensible
  if it is neither $x/y$-justified nor $x/y$-overruled.
\end{definition}

For any program $P$, the least fixpoint exists by the Knaster-Tarski fixpoint
theorem~\cite{Tar55:Fixpoint,Bir67:LatticeTheory}, because $F_{P,x/y}$
is monotone. It can be constructed by transfinite induction as
follows:
$$\begin{array}{llll}
  J^0_{x/y} & \!\!\!=\!\!\! & \emptyset \\
  J^{\alpha+1}_{x/y} & \!\!\!=\!\!\! & F_{P, x/y}(J^\alpha_{x/y}) & 
  \mbox{for $\alpha \! + \! 1$ a successor ordinal} \\
  J^\lambda_{x/y} & \!\!\!=\!\!\! & \bigcup_{\alpha < \lambda}J^\alpha_{x/y} &
  \mbox{for $\lambda$ a limit ordinal} \\
\end{array}$$
Then there exists a least ordinal $\lambda_0$ such that
$F_{x/y}(J^{\lambda_0}_{x/y}) = J^{\lambda_0}_{x/y} = J_{x/y}$.

\section{Relationships of Notions of Justifiability}
\label{sec:relationship}

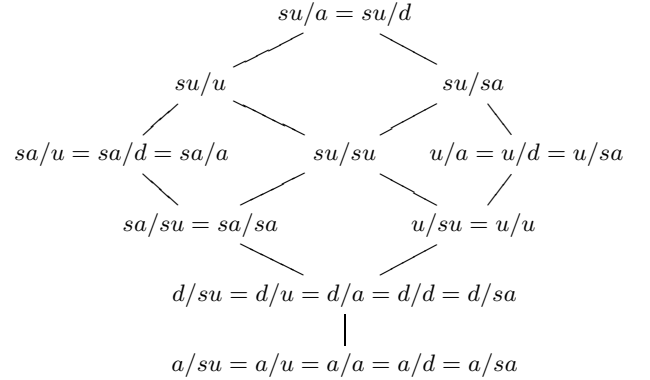
\begin{figure}[htbp]
  \begin{center}\footnotesize
    $\xymatrix@C=-5em@R=3ex{
      && su/a = su/d \\
      & su/u \ar@{-}[ru] && su/sa \ar@{-}[lu] \\
      sa/u = sa/d = sa/a
      \ar@{-}[ru] && 
      su/su \ar@{-}[lu] \ar@{-}[ru] &&
      u/a = u/d = u/sa \ar@{-}[lu] \\
      & sa/su = sa/sa \ar@{-}[lu] \ar@{-}[ru] && 
      u/su = u/u \ar@{-}[lu] \ar@{-}[ru] \\
      && d/su = d/u = d/a = d/d = d/sa
      \ar@{-}[lu] \ar@{-}[ru] \\
      && a/su = a/u = a/a = a/d = a/sa
      \ar@{-}[u] \\
      }$
    \caption{Hierarchy of Notions of Justifiability}
    \label{fig:hierarchy}
  \end{center}
\end{figure}

This section is devoted to an analysis of the relationship 
between the different notions of justifiability, leading to
a hierarchy of notions of justifiability illustrated in 
Figure~\ref{fig:hierarchy}.

First of all, it is easy to see that the least fixpoint increases if
we weaken the attacks, or strengthen the defence.
 
\begin{proposition}
  \label{prop:subset-just}
  Let $x' \subseteq x$, $y \subseteq y'$ be notions of attack, then
  $J_{x/y} \subseteq J_{x'/y'}$.
\end{proposition}

Theorem~\ref{thm:strong-nonstrong} states that it does not make a
difference if we allow only the strong version of the defence. This is
because an argument need not defend itself on its own, but it may rely
on other arguments to defend it.

We only give a formal proof for the first theorem; the proofs for the
other theorems are similar, and we provide an intuitive informal
explanation instead.
 
\begin{theorem}
  \label{thm:strong-nonstrong}
  Let $x$ and $y$ be
  notions of attack such that \mbox{$x \supseteq \mathit{undercuts}$}, 
  and let $sy = y - \mathit{undercuts}^{-1}$.  Then
  $J_{x/y} = J_{x/sy}$.
\end{theorem}

\begin{proof} 
  Informally, every $x$-attack $B$ to an $x/y$-justified argument $A$ is
  $y$-defended by some $x/sy$-justified argument $C$ (by induction). 
  Now if $C$ was {\em not} a $sy$-attack, then it is undercut by $B$,
  and because $x \supseteq \mathit{undercuts}$ and 
  $C$ is justified, there exists a {\em strong} defence
  for $C$ against $B$, which is also a defence of the original
  argument $A$ against $C$.\\[2ex]
  The formal proof is by transfinite induction.
  By Proposition~\ref{prop:subset-just}, we have $J_{x/sy} \subseteq J_{x/y}$.
  We prove the inverse inclusion by showing that
  for all ordinals $\alpha$:
  $J_{x/y}^\alpha \subseteq J_{x/sy}^\alpha$,
  by transfinite induction on $\alpha$.\\[1ex]
  {\it Base case} $\alpha=0$:
  $J_{x/y} = \emptyset = J_{x/sy}$.\\[1ex]
  {\it Successor ordinal } $\alpha \leadsto \alpha+1$:
  Let $A \in J_{x/y}^{\alpha+1}$, and $(B,A) \in x$.
  By definition, there exists $C \in J_{x/y}^\alpha$ such that
  $(C,B) \in y$. By induction hypothesis, $C \in J_{x/sy}^\alpha$.

  If $B$ does not undercut $C$, then we are done. 
  If, however, $B$ undercuts $C$, then because 
  $C \in J_{x/sy}^\alpha$, and $\mathit{undercuts} \subseteq x$,
  there exists 
  $D \in J_{x/sy}^{\alpha_0}(\emptyset) (\alpha_0 < \alpha)$ such that
  $(D,B) \in sy$. It follows that $A \in J_{x/sy}^{\alpha+1}$.\\[1ex]
  {\it Limit ordinal $\lambda$: }
  Assume $J_{x/y}^\alpha \subseteq J_{x/sy}^\alpha$ for all $\alpha < \lambda$.
  Then 
  $J_{x/y}^\lambda = \bigcup_{\alpha<\lambda}J_{x/y}^\alpha \subseteq
  \bigcup_{\alpha<\lambda}J_{x/sy}^\alpha = J_{x/sy}^\lambda$
\end{proof}

In particular, the previous Theorem states that undercut and strong
undercut are equivalent as a defence, as are attack and strong attack.
This may be useful in an implementation, where we may use the stronger
notion of defence without changing the semantics, thereby decreasing
the number of arguments to be checked.
The following Corollary shows that because defeat lies between attack
and strong attack, it is equivalent to both as a defence.

\begin{corollary}
  \label{cor:xa-xsa-just}
  Let $x$ be a notion of attack such that 
  \mbox{$x \supseteq \mathit{undercuts}$}. 
  Then $J_{x/a} = J_{x/d} = J_{x/sa}$.
\end{corollary}

\begin{proof}
  With Proposition~\ref{prop:subset-just} and
  Theorem~\ref{thm:strong-nonstrong}, we have
  $J_{x/a} \subseteq J_{x/d} \subseteq J_{x/sa} = J_{x/a}$.
\end{proof}

\begin{theorem}
  \label{thm:xu-xa-just}
  Let $x$ be a notion of attack such that
  \mbox{$x \supseteq \mbox{\it strongly attacks}$}. Then
  $J_{x/u} = J_{x/d} = J_{x/a}$.
\end{theorem}

\begin{proof}
  Every $x$-attack $B$ to a $x/a$-justified argument $a$
  is attacked by some
  $x/u$-justified argument $C$ (by induction). 
  If $C$ is a rebut, but not an
  undercut, then because $B$ strongly attacks $C$, and because
  $x \supseteq \mbox{\it strongly attacks}$,
  there must have been an argument defending $C$ by
  undercutting $B$, thereby also defending $A$ against $B$.

  The statement for \textit{defeats} follows in a similar way to
  Corollary~\ref{cor:xa-xsa-just}. 

\end{proof}

\begin{theorem}
  \label{thm:sasu-sasa-just}
  $J_{sa/su} = J_{sa/sa}$
\end{theorem}

The proof is similar to Theorem~\ref{thm:xu-xa-just}.

\begin{theorem}
  \label{thm:sua-sud-just}
  $J_{su/a} = J_{su/d}$
\end{theorem}

\begin{proof}
  Every strong undercut $B$ to a $su/a$-justified argument $A$ is attacked
  by some $su/d$-justified argument $C$ (by induction).
  If $C$ does not defeat $A$, then there is some argument $D$
  defending $C$ by defeating $B$, thereby also defending $A$ against $B$.
\end{proof}

We will now present some example programs which distinguish
various notions of justifiability.

\begin{figure}
\footnotesize
\begin{center}
\begin{tabular}{l|l|l}
    $\begin{array}{rcl}
        \multicolumn{3}{c}{P_1 =} \\
      p & \gets & not~q \\
      q & \gets & not~p \\
\\
\\
\\
    \end{array}$
&
    $\begin{array}{rcl}
\multicolumn{3}{c}{P_2 =} \\
      p & \gets & not~q \\
      q & \gets & not~p \\
      \neg p \\
\\
\\
    \end{array}$
&
    $\begin{array}{rcl}
\multicolumn{3}{c}{P_3 =}\\
      p & \gets & not~q \\
      q & \gets & not~r \\
      r & \gets & not~s \\
      s & \gets & not~p \\
      \neg p \\
    \end{array}$
\\\hline
    $\begin{array}{rcl}
\multicolumn{3}{c}{P_4 =} \\
      p & \gets & not~q \\
      q & \gets & not~p \\
      r & \gets & not~p \\ 
\\
    \end{array}$
&
    $\begin{array}{rcl}
\multicolumn{3}{c}{P_5 =} \\
      p & \gets & not~\neg p \\
      \neg p \\
\\
\\
    \end{array}$
&
    $\begin{array}{rcl}
      \multicolumn{3}{c}{P_6 =} \\
      \neg p & \gets & not~q \\
      \neg q & \gets & not~p \\
      p \\
      q \\
    \end{array}$
\end{tabular}
\end{center}
\caption{Examples}
\label{fig:ex}
\end{figure}
\begin{example}\hspace{0cm}
  \label{ex:loop}
Consider $P_1$ in Figure \ref{fig:ex}.
  For any notion of attack $x$, we have
  $J_{su/x} = J_{sa/x} = \{ [p \gets not~q], [q \gets not~p] \},$
  because there is no strong undercut or strong attack to any
  of the arguments.
  However, 
  $J_{a/x} = J_{d/x} = J_{u/x} = \emptyset,$
  because every argument is undercut (and therefore defeated and
  attacked).
\end{example}

\begin{example}\hspace{0em}
  \label{ex:coherence}
Consider $P_2$ in Figure \ref{fig:ex}.
  Let $x$ be a notion of attack. Then 
  $J_{d/x} = J_{a/x} = \emptyset,$
  because every argument is defeated (hence attacked).
  $J_{sa/su} = J_{sa/sa} = \{ [q \gets not~p] \},$
  because $[q \gets not~p]$ is the only argument which is not
  strongly attacked, but it does not strongly attack any other
  argument.
  $J_{u/su} = J_{u/u} = \{ [\neg p] \},$
  because there is no undercut to $[\neg p]$, but 
  $[\neg p]$ does not undercut any other argument.
  $J_{u/a} = \{ [\neg p], [q \gets not~p] \},$
  because there is no undercut to $[\neg p]$, and the undercut
  $[p \gets not~p]$ to $[q \gets not~p]$ is attacked by 
  $[\neg p]$. We also have
  $J_{sa/u} = \{ [\neg p], [q \gets not~p] \},$
  because $[q \gets not~p]$ is not strongly attacked, and the strong
  attack $[p \gets not~q]$ on $[\neg p]$ is undercut by
  $[q \gets not~p]$. 
\end{example}

\begin{example}\hspace{0cm}
  \label{ex:coherence-big}
Consider $P_3$ in Figure \ref{fig:ex}.
  Let $x$ be a notion of attack.
  Then $J_{sa/x} = \emptyset,$
  because every argument is strongly attacked.

  $J_{su/u} = J_{su/su} = \{ [\neg p] \},$
  because all arguments except $[\neg p]$ are
  strongly undercut, but $[\neg p]$ does not undercut any argument.
  And 
  $J_{u/a} = J_{su/sa} = J_{su/a} = 
    \{ [\neg p], [q \gets not~r], [s \gets not~p] \}.$
\end{example}

\begin{example}\hspace{0cm}
  \label{ex:loop-plus}
Consider $P_4$ in Figure \ref{fig:ex}.
  Let $x$ be a notion of attack. Then
  $J_{u/x} = J_{d/x} = J_{a/x} = \emptyset,$
  because every argument is undercut.
  $J_{su/su} = J_{su/sa} = J_{sa/su} = J_{sa/sa} = 
    \{ [p \gets not~q], [q \gets not~p] \}$
  In this case, the strong attacks are precisely the strong undercuts;
  The argument $[r \gets not~p]$ is not justified, because the 
  strong undercut $[p \gets not~q]$ is undercut, but not strongly 
  undercut, by $[q \gets not~p]$.
  $J_{su/u} = J_{su/a} = J_{sa/u} = J_{sa/a} = 
    \{ [p \gets not~q], [q \gets not~p], [r \gets not~p] \}$
  Again, undercuts and attacks, and strong undercuts and strong
  attacks, coincide; but now $[r \gets not~p]$ is justified,
  because non-strong undercuts are allowed as defence.
\end{example}

\begin{example}\hspace{0cm}
  \label{ex:defeat}
Consider $P_5$ in Figure \ref{fig:ex}.
  Then $J_{a/x} = \emptyset$, because both arguments attack each
  other, while $J_{d/x} = \{[\neg p]\}$, because $[\neg p]$ defeats
  $[p \gets not~\neg p]$, but not vice versa.
\end{example}

\begin{example}\hspace{0cm}
  \label{ex:loop-neg}
Consider $P_6$ in Figure \ref{fig:ex}.
  Let $x$ be a notion of attack. Then
  $J_{sa/x} = J_{d/x} = J_{a/x} = \emptyset,$
  because every argument is strongly attacked (hence defeated and attacked),
  while
  $J_{u/x} = J_{su/x} = \{ [p], [q] \}.$
\end{example}

\begin{theorem}
  \label{thm:hierarchy}
  The notions of justifiability are ordered (by set inclusion)
  according to the Hasse diagram in Figure~\ref{fig:hierarchy}.
\end{theorem}

By definition, Dung's grounded argumentation semantics~%
\cite{Dun93:ArgumentExplicit} is exactly $a/u$-justifiability, 
while Prakken and Sartor's semantics~\cite{PS97:Argument},
if we disregard priorities, amounts to $d/su$-justifiability.
As corollaries to Theorem~\ref{thm:hierarchy}, we obtain relationships of
these semantics to the other notions of justifiability.

\begin{corollary}
  Let $J_{Dung}$ be the set of justified arguments according to 
  Dung's grounded argumentation semantics~\cite{Dun93:ArgumentExplicit}.
  Then $J_{Dung}=J_{a/su}=J_{a/u}=J_{a/a}=J_{a/d}=J_{a/sa}$ and
  $J_{Dung} \subseteq J_{x/y}$ for all notions of attack $x$ and $y$.
\end{corollary}

\begin{corollary}
  Let $J_{PS}$ be the set of justified arguments according to 
  Prakken and Sartor's argumentation semantics~\cite{PS97:Argument},
  where all arguments have the same priority.
  Then $J_{PS}=J_{d/su}=J_{d/u}=J_{d/a}=J_{d/d}=J_{d/sa}$,
  $J_{PS} \subseteq J_{x/y}$ for all notions of attack $x \not= a$ and $y$,
  and $J_{PS} \supseteq J_{a/y}$ for all notions of attack $y$.
\end{corollary}

\begin{remark}\hspace{0cm}
1. The notions of $a/x$-, $d/x$- and $sa/x$-justifiability are very
    sceptical in that a {\em fact} $p$ may not be justified, if there
    is a rule $\neg p \gets B$ (where $not~p \not\in B$) that is not
    $x$-attacked. On the other hand this is useful in terms of
    avoiding inconsistency.

\noindent
2.  $sx/y$-justifiability is  very credulous,
    because it does not take into account non-strong attacks, so e.g.\
    the program $\{ p \gets not~q, q \gets not~p \}$ has the justified
    arguments $[p \gets not~q]$ and $[q \gets not~p]$.

\end{remark}

\begin{remark}
  One might ask whether any of the semantics in Figure~\ref{fig:hierarchy}
  are equivalent for {\em non-contradictory} programs, i.e.\ programs for
  which there is no literal $L$ such that there exist justified
  arguments for both $L$ and $\neg L$. 
  The answer to this question is no: all the examples above distinguishing
  different notions of justifiability involve only non-contradictory 
  programs.

  In particular, even for non-contradictory programs,
  Dung's and Prakken and Sartor's semantics differ, and
  both differ from $u/a$-justifiability, which will be shown equivalent to
  the well-founded semantics WFSX~\cite{PA92:WFSX,AP96:WFSX} in the
  following section.
\end{remark}

\section{Well-founded semantics}
\label{sec:wfsx}

We recollect the definition of the well-founded semantics for 
extended logic programs, WFSX. We use the definition of~\cite{ADP95:LPsystem},
because it is closer to our definition of argumentation semantics than
the original definition of~\cite{PA92:WFSX,AP96:WFSX}.

\begin{definition} 
  The set of all objective literals of a program $P$ is 
  called the {\em Herbrand base} of $P$ and denoted by $\herbrand(P)$.
  A {\em pseudo-interpretation} of a program $P$ is a set
  $T \cup not~F$ where $T$ and $F$ are subsets of $\herbrand(P)$.
  An {\em interpretation} is a pseudo-interpretation where the sets $T$ and $F$
  are disjoint. An interpretation is called {\em two-valued} if
  $T \cup F = \herbrand(P)$.
\end{definition}

\begin{definition}
  Let $P$ be an extended logic program, $I$ an interpretation,
  and let $P'$ (resp.~$I'$) be obtained from $P$ (resp.~$I$) by 
  replacing every literal $\neg A$ by a new atom, say $\neg\_A$.
  The GL-transformation $\frac{P'}{I'}$ is the program obtained
  from $P'$ by removing all rules containing a default literal
  $not~A$ such that $A \in I'$, and then removing all remaining
  default literals from $P'$, obtaining a definite program $P''$. 
  Let $J$ be the least model of $P''$. $\Gamma_P I$ is obtained from
  $J$ by replacing the introduced atoms $\neg\_A$ by $\neg A$.
\end{definition}

\begin{definition}
  The {\em semi-normal} version of a program $P$ is the program $P_s$
  obtained from $P$ by replacing every rule $L \gets Body$ in $P$
  by the rule $L \gets not~\neg L, Body$.
\end{definition}

If the program $P$ is clear from the context, we write $\Gamma I$ for
$\Gamma_P I$ and $\Gamma_s I$ for $\Gamma_{P_s} I$.

\begin{definition}
  Let $P$ be a program whose least fixpoint of $\Gamma \Gamma_s$ is $T$.
  Then the {\em paraconsistent well-founded model of $P$} is
  the pseudo-interpretation
  $WFM_p(P) = T \cup not~(\herbrand(P) - \Gamma_s T)$.
  If $WFM_p(P)$ is an interpretation, then
  $P$ is called {\em non-contradictory},
  and $WFM_p(P)$ is the {\em well-founded model of $P$},
  denoted by $WFM(P)$.
\end{definition}
The paraconsistent well-founded model can by defined iteratively by
the transfinite sequence $\{I_\alpha\}$:

  \begin{tabular}{llll}
    $I_0$ & = & $\emptyset$ \\
    $I_{\alpha+1}$ & = & $\Gamma \Gamma_s I_\alpha$ &
    for successor ordinal $\alpha+1$ \\
    $I_\lambda$ & = & $\bigcup_{\alpha < \lambda} I_\alpha$ & 
    for limit ordinal $\lambda$ \\
  \end{tabular}

There exists a smallest ordinal $\lambda_0$ such that $I_{\lambda_0}$
is the least fixpoint of $\Gamma \Gamma_s$, and 
$WFM_p(P) = I_{\lambda_0} \cup not~(\herbrand(P) - \Gamma_s I_{\lambda_0})$.

\section{Equivalence of argumentation semantics and WFSX}
\label{sec:wfsxarg}

In this section, we will show that the argumentation semantics $J_{u/a}$ and 
the well-founded model coincide.
That is, the conclusions of justified arguments
are exactly the objective literals which are true in the well-founded model;
and those objective literals all of whose arguments are overruled are exactly
the literals which are false in the well-founded model.
The result holds also for contradictory programs under the
{\em paraconsistent} well-founded semantics.
This is important, because it shows that contradictions in the
argumentation semantics are precisely the contradictions under
the well-founded semantics, and allows the application of 
contradiction removal (or avoidance) methods to the argumentation semantics.
Because for non-contradictory programs, 
the well-founded semantics coincides with the paraconsistent
well-founded semantics~\cite{ADP95:LPsystem}, 
we obtain as a corollary that argumentation semantics and well-founded
semantics coincide for non-contradictory programs.

In order to compare the argumentation semantics with the well-founded
semantics, we define the set of literals which are a
consequence of the argumentation semantics.

\begin{definition}
  $A(P) = T~\cup~not~F$, where \\
  $T = \{ L ~|$ there is a justified argument for $L \}$ and \\
  $F = \{ L ~|$ all arguments for $L$ are overruled $\}$.
\end{definition}

The following Proposition shows a precise connection between arguments and
consequences of a program $\frac{P}{I}$.
\begin{proposition}
  \label{prop:model-arg}
  Let $I$ be a two-valued interpretation.
  \begin{enumerate}
  \item \label{prop:model-arg:u}
    $L \in \Gamma(I)$ iff $\exists$ argument $A$ with conclusion $L$
    such that $\assm(A) \subseteq I$.
  \item \label{prop:model-arg:a}
    $L \in \Gamma_s(I)$ iff $\exists$ argument $A$ with conclusion $L$
    such that $\assm(A) \subseteq I$ and
    $\neg\conc(A) \cap I = \emptyset$.
  \item \label{prop:model-arg:u-neg}
    $L \not\in \Gamma(I)$ iff $\forall$ arguments $A$ with conclusion $L$,
    $\assm(A) \cap I \not= \emptyset$.
  \item \label{prop:model-arg:a-neg}
    $L \not\in \Gamma_s(I)$ iff $\forall$ arguments $A$ with 
    conclusion $L$, $\assm(A) \cap I \not= \emptyset$
    or $\neg\conc(A) \cap I \not= \emptyset$.
  \end{enumerate}
\end{proposition}
\begin{proof}
  See Appendix.
\end{proof}

\begin{theorem}
  \label{thm:wfsxarg}
  Let $P$ be an extended logic program.
  Then $WFM_p(P) = A(P)$.
\end{theorem}

\begin{proof}\raggedright
  First, note that $A$ undercuts $B$ iff 
  $\exists~L$ s.t.\ $not~L \in \assm(A)$ and $L \in \conc(B)$;
  and $A$ rebuts $B$ iff $\exists~L \in \conc(A) \cap \neg\conc(B)$.\\[2ex]
  We show that for all ordinals $\alpha$, $I_\alpha = A_\alpha$,
  by transfinite induction on $\alpha$.\\[0.5ex]
  \textit{Base case $\alpha=0$:}
  $I_\alpha = \emptyset = A_\alpha$
  \\[0.5ex]
  \textit{Successor ordinal $\alpha \leadsto \alpha+1$}:\\[0.5ex]
  $L \in I_{\alpha+1}$ \\ 
  \mycenter{iff (Def.\ of $I_{\alpha+1}$)} \\
  $L \in \Gamma \Gamma_s I_\alpha$  \\
  \mycenter{iff (Prop.~\ref{prop:model-arg}(\ref{prop:model-arg:u}))} \\
  $\exists$~argument $A$ for $L$ such that
  $\assm(A) \subseteq \Gamma_s I_\alpha$ \\
  \mycenter{iff (Def.\ of $\subseteq$, and $\Gamma_s I_\alpha$ 
    is two-valued)} \\
  $\exists$~argument $A$ for $L$ such that
  $\forall~not~L \in \assm(A), L \not\in \Gamma_s I_\alpha$ \\ 
  \mycenter{iff (Prop.~\ref{prop:model-arg}(\ref{prop:model-arg:a-neg}))} \\
  $\exists$~argument $A$ for $L$ such that $\forall~not~L \in \assm(A)$, 
  for any argument $B$ for $L$,
  (~$\exists~not~L' \in \assm(B) ~s.t.\ L' \in I_\alpha$
    or $\exists~L'' \in \conc(B) ~s.t.\ \neg L'' \in I_\alpha$ ) \\
  \mycenter{iff (Induction hypothesis)} \\
  $\exists$~argument $A$ for $L$ such that $\forall~not~L \in \assm(A)$, 
  for any argument $B$ for $L$,
  (~$\exists~not~L' \in \assm(B) ~s.t.\ \exists$
    argument $C \in J_\alpha$ for $L'$,
    or $\exists~L'' \in \conc(B) ~s.t.\ \exists$~
    argument $C \in J_\alpha$ for $\neg L''$) \\
  \mycenter{iff (Def.\ of undercut and rebut)} \\
  $\exists$~argument $A$ for $L$ such that for any undercut $B$ to $A$,
  (~$\exists$~argument $C \in J_\alpha$ s.t.\
    $C$ undercuts $B$,
    or $\exists$~argument $C \in J_\alpha$ s.t.\
    $C$ rebuts $B$) \\
  \mycenter{iff} \\
  $\exists$~argument $A$ for $L$ such that for any undercut $B$ to $A$,
  $\exists$~argument $C \in J_\alpha$ s.t.\ $C$ attacks $B$ \\
  \mycenter{iff (Def.\ of $J_{\alpha+1}$)} \\
  $\exists$~argument $A \in J_{\alpha+1}$ for $L$ \\
  \mycenter{iff  (Def.\ of $A_{\alpha+1}$)} \\
  $L \in A_{\alpha+1}$\\[0.5ex]
  \textit{Limit ordinal $\lambda$:}\\
  $I_\lambda = \bigcup_{\alpha < \lambda} I_\alpha$ and
  $A_\lambda = \bigcup_{\alpha < \lambda} A_\alpha$,
  so by induction hypothesis ($I_\alpha = A_\alpha$ for all $\alpha < \lambda$),
  $I_\lambda = A_\lambda$.\\[2ex]
  Now, we show that a literal $not~L$ is in the well-founded semantics
  iff every argument for $L$ is overruled.\\[0.5ex]
  $not~L \in WFM_p(P)$ \\
  \mycenter{iff (Def.\ of $WFM_p(P)$)} \\
  $L \not\in \Gamma_s I$ \\
  \mycenter{iff (Prop.~\ref{prop:model-arg}(\ref{prop:model-arg:a-neg}))} \\
  for all arguments $A$ for $L$,
  (~$\exists~not~L' \in \assm(A) ~s.t.\ L' \in I$, or
    $\exists~L'' \in \conc(A) ~s.t.\ \neg L'' \in I$ ) \\
  \mycenter{iff (I=A)} \\
  for all arguments $A$ for $L$,
  (~$\exists~not~L' \in \assm(A) ~s.t.\ 
    \exists$~argument $B \in J$ for $L'$, or
    $\exists~L'' \in \conc(A) ~s.t.\ 
    \exists$~argument $B \in J$ for $\neg L''$ ) \\
  \mycenter{iff (Def.\ of undercut and rebut)} \\
  for all arguments $A$ for $L$,
  (~$\exists$~argument $B \in J$ s.t.\ $B$ undercuts $A$, or
    $\exists$~argument $B \in J$ s.t.\ $B$ rebuts $A$ ) \\
  \mycenter{iff} \\
  every argument for $L$ is attacked by a justified argument in $J$ \\
  \mycenter{iff (Def.\ of overruled)} \\
  every argument for $L$ is overruled \\
  \mycenter{iff (Def.\ of $A(P)$)} \\
  $not~L \in A(P)$
\end{proof}

\begin{corollary}
  Let $P$ be a non-contradictory program. Then $WFM(P) = A(P)$.
\end{corollary}

\begin{remark}
  In a similar way, one can show that the $\Gamma$ operator corresponds
  to undercuts, while the $\Gamma_s$ operator corresponds to attacks, and so
  the least fixpoints of $\Gamma\Gamma$, $\Gamma_s\Gamma$, and
  $\Gamma_s\Gamma_s$ correspond to $J_{u/u}$, $J_{a/u}$, and $J_{a/a}$,
  respectively. 
  In~\cite{ADP95:LPsystem}, the least fixpoints of these operators are
  shown to be ordered as 
  $lfp(\Gamma_s\Gamma) \subseteq lfp(\Gamma_s\Gamma_s) \subseteq
   lfp(\Gamma\Gamma_s)$, and 
  $lfp(\Gamma_s\Gamma) \subseteq lfp(\Gamma\Gamma) \subseteq 
   lfp(\Gamma\Gamma_s)$. 
  Because $J_{a/u} = J_{a/a} \subseteq J_{u/u} \subseteq J_{u/a}$ by 
  Theorem~\ref{thm:hierarchy}, we can strengthen this statement
  to $lfp(\Gamma_s\Gamma) = lfp(\Gamma_s\Gamma_s) \subseteq
  lfp(\Gamma\Gamma) \subseteq lfp(\Gamma\Gamma_s)$.
\end{remark}

\section{Proof theory}
\label{sec:proof-theory}

One of the benefits of relating the argumentation semantics $J_{u/a}$ to
WFSX is the existence of an efficient top-down proof procedure for 
WFSX~\cite{ADP95:LPsystem},
which we can use to compute justified arguments in $J_{u/a}$.
On the other hand, {\em dialectical} proof theories, based on dialogue trees,
have been defined for a variety of argumentation semantics
\cite{PS97:Argument,JV99:DialecticSemantics,KT99:ComputingArgumentation}.
In this section, we present a sound and complete dialectical proof theory for
the least fixpoint argumentation semantics $J_{x/y}$ for any
notions of attack $x$ and $y$. Our presentation closely 
follows~\cite{PS97:Argument}.
As a further consequence, we obtain an equivalence of the proof theory
for WFSX and the dialectical proof theory for arguments.

\begin{definition}
  An {\em $x/y$-dialogue} is a finite nonempty sequence of 
  moves $move_i = (\mathit{Player}_i,Arg_i) (i > 0)$, such that
  \begin{enumerate}
  \item $\mathit{Player}_i = P$ iff $i$ is odd; and 
    $\mathit{Player}_i = O$ iff $i$ is even.
  \item If $\mathit{Player}_i = \mathit{Player}_j$ and $i \not= j$,
    then $Arg_i \not= Arg_j$.
  \item If $\mathit{Player}_i = P$ and $i > 1$, then
    $Arg_i$ is a minimal argument such that $(Arg_i,Arg_{i-1}) \in y$.
  \item If $\mathit{Player}_i = O$, then $(Arg_i,Arg_{i-1}) \in x$.
  \end{enumerate}
\end{definition}

The first condition states that the players $P$ (Proponent) and
$O$ (Opponent) take turns, and $P$ starts.
The second condition prevents the proponent from repeating a move.
The third and fourth conditions state that both players have to 
attack the other player's last move, where the opponent is allowed
to use the notion of attack $x$, while the proponent may use $y$ to 
defend its arguments.

\begin{definition}
  An {\em $x/y$-dialogue tree} is a tree of moves such that
  every branch is a dialogue, and for all moves $move_i = (P, Arg_i)$,
  the children of $move_i$ are all those moves $(O,Arg_j)$ such that
  $(Arg_j,Arg_i) \in x$.
\end{definition}

\begin{definition}
  A player {\em wins an $x/y$-dialogue} iff the other player cannot move.
  A player {\em wins an $x/y$-dialogue tree} iff it wins all branches
  of the tree. An $x/y$-dialogue tree which is won by the proponent
  is called a {\em winning $x/y$-dialogue tree}.
  An argument $A$ is {\em provably $x/y$-justified} iff there exists a
  $x/y$-tree with $A$ as its root, and won by the proponent.
  A literal $L$ is a {\em provably justified conclusion} iff
  it is a conclusion of a provably $x/y$-justified argument.
  The {\em height} of a dialogue tree is $0$ if it consists only of
  the root, and otherwise $height(t) = \bigcup height(t_i)+1$ where 
  $t_i$ are the trees rooted at the grandchildren of $t$.
\end{definition}

We show that the proof theory of $x/y$-dialogue trees is sound and
complete for any notions of attack $x$ and $y$.

\begin{theorem}
  \label{thm:sound-complete}
  An argument is provably $x/y$-justified iff it is $x/y$-justified.
\end{theorem}

\begin{proof}
  ``If''-direction.
  We show by transfinite induction: If $A \in J^{\alpha}_{x/y}$, then 
  there exists a winning $x/y$-dialogue tree of height $< \alpha$ 
  for $A$.\\[1ex]
  {\em Base case $\alpha=0$:}\\
  Then there exists no argument $B$ such that $(B,A) \in x$,
  and so $A$ is a winning $x/y$-dialogue tree for $A$ of height $0$.\\[0.5ex]
  {\em Successor ordinal $\alpha+1$:}\\
  If $A \in J_{x/y}^{\alpha+1}$, then for any $B_i$ such that $(B_i,A) \in x$
  there exists a $C_i \in J_{x/y}^{\alpha}$ such that $(C_i,B_i) \in y$.
  By induction hypothesis, there exist winning $x/y$-dialogue trees
  for the $C_i$. Thus, we have a winning tree rooted for $A$,
  with children $B_i$, whose children are the winning trees for $C_i$.\\[0.5ex]
  {\em Limit ordinal $\lambda$:}\\
  If $A \in J_{x/y}^{\lambda}$, then there exists an $\alpha < \lambda$
  such that $A \in J_{x/y}^{\alpha}$; by induction hypothesis,
  there exists a winning $x/y$-dialogue tree of height $\alpha$ for $A$.\\[1ex]
  ``Only-if''-direction.
  We prove by transfinite induction: 
  If there exists a winning tree of height $\alpha$ for $A$,
  then $A \in J_{x/y}^{\alpha}$.\\[1ex]
  {\em Base case $\alpha=0$:}\\
  Then there are no arguments $B$ such that $(B,A) \in x$, and so
  $A \in J_{x/y}^0$.\\[0.5ex]
  {\em Successor ordinal $\alpha+1$:}\\
  Let $T$ be a tree with root $A$, whose children are $B_i$,
  and the children of $B_i$ are winning trees rooted at $C_i$.
  By induction hypothesis, $C_i \in J_{x/y}^{\alpha}$.
  Because the $B_i$ are all those arguments such that $(B_i,A) \in x$,
  then $A$ is defended against each $B_i$ by $C_i$, and so
  $A \in J_{x/y}^{\alpha+1}$.
\end{proof}

As a corollary, we can relate the proof theory of WFSX and the 
$u/a$-proof theory.

\begin{corollary}
  $L$ is a provably $u/a$-justified conclusion iff
  there exists a successful T-tree~\cite{AP96:WFSX} for $L$.
\end{corollary}

\begin{proof}
  Follows from the fact that $u/a$-dialogue trees are sound and complete
  for $u/a$-justifiability (Theorem~\ref{thm:sound-complete}),
  that $T$-trees are sound and complete for WFSX~\cite{AP96:WFSX},
  and that $u/a$-justifiability and WFSX are equivalent
  (Theorem~\ref{thm:wfsxarg}).
\end{proof}

\section{Conclusion and Further Work}
\label{sec:conclusion}

We have identified various notions of attack for extended logic
programs. Based on these notions of attack, we defined notions of
acceptability and least fixpoint semantics. These fixpoint semantics
were related by establishing a lattice of justified arguments, based
on set inclusion.  We identified an argumentation semantics $J_{u/a}$
equal to the well-founded semantics for logic programs with explicit
negation, $WFSX$~\cite{AP96:WFSX}, and established that $J_{Dung}
\subseteq J_{PS} \subseteq J_{u/a} = WFSX$, where $J_{Dung}$ and
$J_{PS}$ are the least fixpoint argumentation semantics of
Dung~\cite{Dun93:ArgumentExplicit} and Prakken and
Sartor~\cite{PS97:Argument}.  We have defined a dialectical proof
theory for argumentation.  For all notions of justified arguments
introduced, we prove that the proof theory is sound and complete
wrt. the corresponding fixpoint argumentation semantics. In
particular, we showthe equivalence of successful T-trees
~\cite{AP96:WFSX} in WFSX to provably $u/a$ justified arguments.

Finally, it remains to be seen whether a variation in the notion of
attack yields interesting variations of alternative argumentation
semantics for extended logic programs such as preferred extensions or
stable extensions~\cite{Dun93:ArgumentExplicit}.  It is also an open
question how the hierarchy changes when priorities are added as
defined in \cite{PS97:Argument, Vre97:AbstractArgumentationSystems}.

\noindent {\bf Acknowledgement}
This work has been supported by EPSRC grant GRM88433.

\bibliographystyle{plain}

\newpage
\appendix

\section*{Appendix}

\begin{proof}[Proof of Proposition~\ref{prop:model-arg}]\hspace{0em}
  \raggedright
  \begin{enumerate}
  \item ``If''-direction:
    Induction on the length $n$ of the derivation of $L \in \Gamma(I)$.\\[0.5ex]
    \textit{Base case}: $n = 1$:\\
    Then there exists a rule $L \gets not~L_1,\ldots,not~L_n$ in $P$ s.t.\
    $L_1,\ldots,L_n \not\in I$,
    and $[L \gets not~L_1,\ldots,not~L_n]$ is an argument for $L$ 
    whose assumptions are contained in $I$.\\[0.5ex]
    \textit{Induction step}: $n \leadsto n+1$:\\
    Let $L \in \Gamma^{n+1}(I)$. Then there exists a rule 
    $r = L \gets L_1,\ldots,L_n,not~L'_1,\ldots,L'_m$ in $P$ s.t.\
    $L_i \in \Gamma^n(I)$, and $L'_i \not\in I$.
    By induction hypothesis, there exists arguments $A_1,\ldots,A_n$ for
    $L_1,\ldots,L_n$ with $\assm(A_i) \subseteq I$.
    Then $A = [r] \cdot A_1 \cdots A_n$ is an argument for $L$
    such that $\assm(A) \subseteq I$.\\[1ex]
    ``Only-if'' direction: Induction on the length of the argument.\\[0.5ex]
    \textit{Base case}: $n = 1$:\\
    Then $A = [L \gets not~L_1,\ldots,not~L_n]$, and 
    $L_1,\ldots,L_n \not\in I$. Then $L \gets \in \frac{P}{I}$,
    and $L \in \Gamma^1(I)$.\\[0.5ex]
    \textit{Induction step}: $n \leadsto n+1$:\\
    Let $A = [L \gets L_1,\ldots,L_n,not~L'_1,\ldots,not~L'_m;r_2,\ldots,r_n]$
    be an argument s.t.\ $\assm(A) \subseteq I$.
    $A$ contains subarguments $A_1,\ldots,A_n$ for $L_1,\ldots,L_n$,
    with $\assm(A_i) \subseteq I$. 
    Because $L'_1,\ldots,L'_m \not\in I$, then
    $L \gets L_1,\ldots,L_n \in \frac{P}{I}$.
    By induction hypothesis, $L_i \in \Gamma(I)$.
    so also $L \in \Gamma(I)$.
  \item ``If''-direction:
    Induction on the length $n$ of the derivation of $L \in \Gamma_s(I)$.\\[0.5ex]
    \textit{Base case}: $n = 1$:\\
    Then there exists a rule $L \gets not~L_1,\ldots,not~L_n$ in $P$ s.t.\
    $\neg L,L_1,\ldots,L_n \not\in I$,
    and $[L \gets not~L_1,\ldots,not~L_n]$ is an argument for $L$ 
    whose assumptions are contained in $I$, and $\neg L \not\in I$.\\[0.5ex]
    \textit{Induction step}: $n \leadsto n+1$:\\
    Let $L \in \Gamma^{n+1}(I)$. Then there exists a rule 
    $r = L \gets L_1,\ldots,L_n,not~L'_1,\ldots,L'm$ in $P$ s.t.\
    $L_i \in \Gamma^n(I)$, $L'_i \not\in I$, and $\neg L \not\in I$.
    By induction hypothesis, there exists arguments $A_1,\ldots,A_n$ for
    $L_1,\ldots,L_n$ with $\assm(A_i) \subseteq I$ and 
    $\neg \conc(A_i) \cap I = \emptyset$.
    Then $A = [r] \cdot A_1 \cdots A_n$ is an argument for $L$
    such that $\assm(A) \subseteq I$, and $\neg \conc(A) \cap I = \emptyset$.\\[1ex]
\newpage
    ``Only-if'' direction: Induction on the length of the argument.\\[0.5ex]
    \textit{Base case}: $n = 1$:\\
    Then $A = [L \gets not~L_1,\ldots,not~L_n]$, and 
    $\neg L,L_1,\ldots,L_n \not\in I$. Then $L \gets \in \frac{P_s}{I}$,
    and $L \in \Gamma^1(I)$.\\[0.5ex]
    \textit{Induction step}: $n \leadsto n+1$:\\
    Let~$A = [L \gets L_1,\ldots,L_n,not~L'_1,\ldots,not~L'_m;\ldots]$
    be an argument s.t.\ $\assm(A) \subseteq I$, and 
    $\neg \conc(A) \cap I = \emptyset$.
    $A$ contains subarguments $A_1,\ldots,A_n$ for $L_1,\ldots,L_n$,
    with $\assm(A_i) \subseteq I$, and $\neg \conc(A_i) \cap I = \emptyset$. 
    Because $L'_1,\ldots,L'_m \not\in I$, and $\neg L \not\in I$, then
    $L \gets L_1,\ldots,L_n \in \frac{P}{I}$.
    By induction hypothesis, $L_i \in \Gamma(I)$.
    so also $L \in \Gamma(I)$.
  \item and 4. follow immediately from \ref{prop:model-arg:u}. and
    \ref{prop:model-arg:a}., because $I$ is two-valued.
  \end{enumerate}
\end{proof}

\end{document}